\documentclass[aps,prl,twocolumn,amsmath,amssymb,floatfix,superscriptaddress]{revtex4-1}
\usepackage{physics,amsmath}
\usepackage{tabularx}
\usepackage{euscript}
\usepackage{epsfig,psfrag,subfigure}
\usepackage{dcolumn}
\usepackage{bm}
\usepackage{setspace}
\usepackage{appendix}
\usepackage{color,soul, xcolor}
\usepackage{amsfonts}
\usepackage{exscale}
\usepackage{multirow}
\usepackage{hyperref}
\usepackage{wrapfig}
\usepackage{lipsum}
\hypersetup{colorlinks=false,linkcolor=black}
\newcommand{\nips}{NiPS$_3$}
\begin{document}
\title{3D Heisenberg universality in the Van der Waals antiferromagnet \nips}

\author{Rajan Plumley}
\affiliation{Department of Physics, Carnegie Mellon University, Pittsburgh, PA, USA}
\affiliation{Stanford Institute for Materials and Energy Sciences, Stanford University, Menlo Park, CA, USA}
\affiliation{Linac Coherent Light Source, SLAC National Accelerator Laboratory, Menlo Park, CA, USA}

\author{Sougata Mardanya}
\affiliation{Department of Physics, Howard University, Washington DC, USA}

\author{Cheng Peng}
\affiliation{Stanford Institute for Materials and Energy Sciences, Stanford University, Menlo Park, CA, USA}

\author{Johannes Nokelainen}
\affiliation{Department of Physics, Howard University, Washington DC, USA}
\affiliation{Department of Physics, Northeastern University, Boston, USA}

\author{Tadesse Assefa}
\affiliation{Stanford Institute for Materials and Energy Sciences, Stanford University, Menlo Park, CA, USA}

\author{Lingjia Shen}
\affiliation{Stanford Institute for Materials and Energy Sciences, Stanford University, Menlo Park, CA, USA}
\affiliation{Linac Coherent Light Source, SLAC National Accelerator Laboratory, Menlo Park, CA, USA}
\affiliation{Division of Synchrotron Radiation Research, Department of Physics, Lund University, Lund, Sweden}

\author{Nicholas Burdet}
\affiliation{Stanford Institute for Materials and Energy Sciences, Stanford University, Menlo Park, CA, USA}
\affiliation{Linac Coherent Light Source, SLAC National Accelerator Laboratory, Menlo Park, CA, USA}

\author{Zach Porter}
\affiliation{Stanford Institute for Materials and Energy Sciences, Stanford University, Menlo Park, CA, USA}
\affiliation{Linac Coherent Light Source, SLAC National Accelerator Laboratory, Menlo Park, CA, USA}

\author{Alexander Petsch}
\affiliation{Stanford Institute for Materials and Energy Sciences, Stanford University, Menlo Park, CA, USA}
\affiliation{Linac Coherent Light Source, SLAC National Accelerator Laboratory, Menlo Park, CA, USA}

\author{Aidan Israelski}
\affiliation{Linac Coherent Light Source, SLAC National Accelerator Laboratory, Menlo Park, CA, USA}

\author{Hongwei Chen}
\affiliation{Stanford Institute for Materials and Energy Sciences, Stanford University, Menlo Park, CA, USA}
\affiliation{Linac Coherent Light Source, SLAC National Accelerator Laboratory, Menlo Park, CA, USA}
\affiliation{Department of Physics, Northeastern University, Boston, USA}

\author{Jun Sik Lee}
\affiliation{Stanford Synchrotron Light Source, SLAC National Accelerator Laboratory, Menlo Park, CA, USA}

\author{Sophie Morley}
\affiliation{Advanced Light Source, Lawrence Berkeley National Laboratory, Berkeley, CA, USA}

\author{Sujoy Roy}
\affiliation{Advanced Light Source, Lawrence Berkeley National Laboratory, Berkeley, CA, USA}

\author{Gilberto Fabbris}
\affiliation{Advanced Photon Source, Argonne National Laboratory, Lemont, Illinois, USA}

\author{Elizabeth Blackburn}
\affiliation{Division of Synchrotron Radiation Research, Department of Physics, Lund University, Lund, Sweden}

\author{Adrian Feiguin}
\affiliation{Department of Physics, Northeastern University, Boston, USA}

\author{Arun Bansil}
\affiliation{Department of Physics, Northeastern University, Boston, USA}

\author{Wei-Sheng Lee}
\affiliation{Stanford Institute for Materials and Energy Sciences, Stanford University, Menlo Park, CA, USA}

\author{Aaron Lindenberg}
\affiliation{Stanford Institute for Materials and Energy Sciences, Stanford University, Menlo Park, CA, USA}
\affiliation{Department of Materials Science and Engineering, Stanford University, Stanford, USA}

\author{Sugata Chowdhury}
\affiliation{Department of Physics, Howard University, Washington DC, USA}

\author{Mike Dunne}
\affiliation{Linac Coherent Light Source, SLAC National Accelerator Laboratory, Menlo Park, CA, USA}

\author{Joshua J. Turner}
\email{joshuat@slac.stanford.edu}
\affiliation{Stanford Institute for Materials and Energy Sciences, Stanford University, Menlo Park, CA, USA}
\affiliation{Linac Coherent Light Source, SLAC National Accelerator Laboratory, Menlo Park, CA, USA}


\maketitle

\section{abstract}
Van der Waals (vdW) magnetic materials are comprised of layers of atomically thin sheets, making them ideal platforms for studying magnetism at the two-dimensional (2D) limit.  These materials are at the center of a host of novel types of experiments, however, there are notably few pathways to directly probe their magnetic structure.  We confirm the magnetic order within a single crystal of \nips~and show it can be accessed with resonant elastic X-ray diffraction along the edge of the vdW planes in a carefully grown crystal by detecting structurally forbidden resonant magnetic X-ray scattering.  We find the magnetic order parameter has a critical exponent of $\beta\sim0.36$, indicating that the magnetism of these vdW crystals is more adequately characterized by the three-dimensional (3D) Heisenberg universality class.  We verify these findings with first-principles density functional theory, Monte-Carlo simulations, and density matrix renormalization group calculations.

\section{Introduction}
Van der Waals (vdW) magnetic materials have emerged as an excellent platform for investigating low-dimensional magnetism, owing to their unique structural characteristics and tunable magnetic properties~\cite{mcguire2020cleavable, burch2018magnetism, park2016opportunities}. Their remarkable sensitivity to external stimuli, such as electric or magnetic fields, strain, and temperature, makes them highly desirable for applications in next-generation spintronics and quantum computing devices~\cite{gibertini2019magnetic}. Furthermore, they offer an ideal platform for investigating fundamental magnetic Hamiltonians in real physical systems~\cite{joy1992magnetism}. Some of these model Hamiltonians can be solved analytically and have led to many discoveries that have guided the field of low-dimensional magnetism, such as the Berezinskii-Kosterlitz-Thouless (BKT) transition \cite{Berezinskii1971,Kosterlitz_1973} in the Heisenberg XY model. As a result, vdW magnetic materials have garnered tremendous attention from both theorists and experimentalists, who seek to explore their fundamental physics and exploit their technological potential~\cite{gong2019two, park2016opportunities}.

The Transition Metal Chalchogenophosphates (TMCPs), include a class of isostructural vdW antiferromagnets (AFMs) which are easy to synthesize and chemically modify, making them exceptional for exploring low dimensional magnetic properties~\cite{grasso2002low}. In addition, the various compounds within this family all exhibit honeycomb magnetic structures, which can often be modeled by principal spin-Hamiltonians~\cite{joy1992magnetism, park2016opportunities}.  For instance, the honeycomb arrangement of Ni ions in \nips~bears semblance to graphene, and its layers are similarly easy to exfoliate.  Similarly to graphene, \nips~has been established as a rich system for exploring strongly correlated dynamics~\cite{park2016opportunities, basnet2021highly, belvin2021exciton, kang2020coherent, bazazzadeh2021magnetoelastic, afanasiev2021controlling, kim2019mott, kim2019suppression}.  However, there are still many open questions related to whether or not the long-range magnetism in \nips~is intrinsically different in its bulk or monolayer forms. It was recently shown that the `zig-zag' AFM order persists when the thickness is reduced down to two layers but disappears altogether in the monolayer limit~\cite{kim2019suppression}, indicating that pure 2D magnetism may not be stable in this system.  Furthermore, the question of the spin-dimensionality in \nips~is still debated at this time, with some studies reporting it as best being described by a two-component spin (XY-like) \cite{afanasiev2021controlling, wildes2015nips3, kang2020coherent, seifert2022ultrafast} and others reporting it as a three-component spin (Heisenberg-like) ~\cite{bazazzadeh2021magnetoelastic, joy1992magnetism, wildes2022magnetic, scheie2023spin}. Without a well-established form and set of parameters for the spin-Hamiltonian, some aspects of the magnetic dynamics in \nips~remain to be fully described. 

%

\begin{figure*}[tb]
    \centering
    \includegraphics[width=\textwidth]{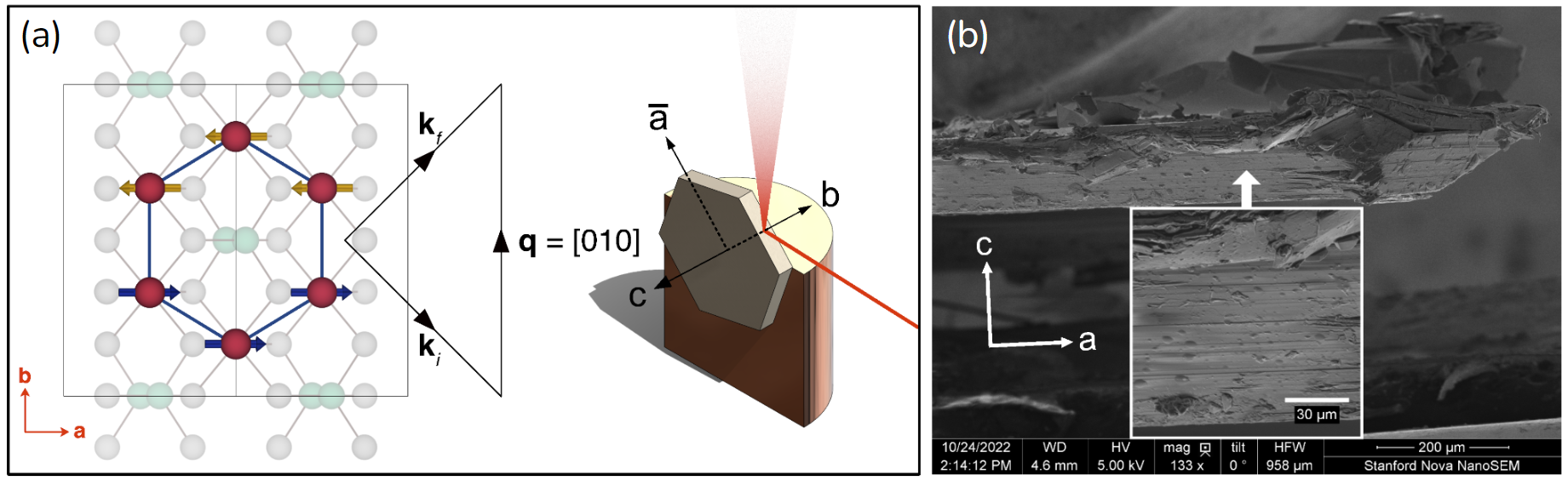}
    \caption{Sample and wide-angle X-ray diffraction geometry. (a) Magnetic unit cell of \nips~viewed along the $c$-axis and X-ray scattering setup schematic showing the orientation of the crystal and copper mount with respect to the beam (shown in red).  The scattering plane is parallel to the $ab$-plane. (b) Scanning electron micrograph of the edge of the \nips~crystal used in the X-ray experiments, viewed from its side profile.  The sample is oriented to be viewed along the $b$-axis in order to locate the optimal location to place the beam to avoid regions where the layers have deformed or delaminated from the bulk crystal.  The beam spot size was approximately 8~\textmu m. Inset: the preferred scattering region where the vdW layers exhibit a smooth surface.}\label{Fig_unitcell}
\end{figure*}


A key current hurdle in studying the detailed magnetic properties of vdW crystals is the scarcity of techniques that can directly characterize the long-range magnetic order within the 2D layer with atomic spatial resolution~\cite{mcguire2020cleavable}. Historically, neutron scattering, which can be used to analyze the magnetic structure and excitations in crystals, has been instrumental in pioneering our understanding of \nips~and related compounds~\cite{wildes2015nips3, wildes2022magnetic, lanccon2016feps3, wildes2006static, wildes2017magnetic, scheie2023spin}, but these are typically conducted on powder samples or large batches of crystal samples fused together. Single vdW crystals are usually small and fragile, which pose inherent limitations for measurements which require large crystals. In this context, synchrotron and X-ray free-electron laser (XFEL) facilities present an opportunity to provide another perspective for analyzing vdW magnets. While typically it requires sufficient surface polishing because of the smaller penetration depth compared to neutrons, X-ray beam spot sizes however can be focused to a few microns such that tiny crystals can be studied. This combined with a strong X-ray cross-section allows thin vdW crystals to be analyzed without averaging over large sample volumes. Additionally, by utilizing their extreme brightness and tunable energy, the ability to probe electronic and magnetic states with elemental specificity is also advantageous. Thus resonant X-ray diffraction is particularly promising for the study of vdW systems, yet remains largely unexplored. 

In this article, we address these challenges by directly studying the properties of the long-range magnetic order in \nips~using resonant X-ray diffraction from a single crystal. We explicitly probe the $d$-states of Ni with an incident X-ray beam energy tuned to the $L$-edge ($2p \rightarrow 3d$) resonance on a system that can not be polished easily due to the delicate nature of these magnetic vdW layers. The in-plane antiferromagnetic order was revealed by orienting the scattering plane of the X-ray beam parallel to the vdW layers, with the beam incident on the natural crystal surface along the edge of the layers. Using theoretical calculations and measuring the critical scaling of the AFM so-called `zig-zag' sublattice order parameter, we find that the magnetic structure surprisingly scales as the 3D Heisenberg universality class. This is in contradiction to the many 2D universality classes conventionally assigned to Van der Waals magnet family~\cite{gibertini2019magnetic, taroni2008universal}. This points to the importance of exploring the nature of these magnetic interactions, such as through magnetic fluctuation studies, where the dimensionality largely determines the role of thermal fluctuations.

\section{Results}
Two resonant X-ray diffraction experiments were carried out at the Ni $L$-edge at 853~eV. These two experiments were performed at the Stanford Synchrotron Radiation Light Source (SSRL) and Advanced Light Source (ALS) facilities. Details about these experimental setups, as well as an additional survey of the magnetic and structural reflections at the Ni $K$-edge ($1s \rightarrow 3d$) at 8.332~keV from the Advanced Photon Source (APS), are discussed in the supplemental materials.

Obtaining the $(0k0)$-reflection from \nips~poses a practical challenge because it requires scattering along the magnetic layer, i.e.~making the scattering plane co-planar with the vdW layers ($ab$-plane). The scattering geometry we employed for all these experiments is shown in Fig.~\ref{Fig_unitcell}a together with an scanning electron microscopy (SEM) image of the crystal Fig.\ref{Fig_unitcell}b used in the X-ray measurements taken with a small beam size at the ALS facility (see experimental section). This was used to explore the preferred scattering region on the crystal edge, where optimal surface regions were identified with a length-scale several times larger than the incident beam spot size. The SEM image shows the morphology of the crystal as seen from its side-profile, where some layers are delaminated from the rest of the bulk crystal in certain regions along the edge. These regions were avoided by marking the position of region with the highest quality and using a small beam size.  

\begin{figure*}[tb]
    \centering
    \includegraphics[width=\textwidth]{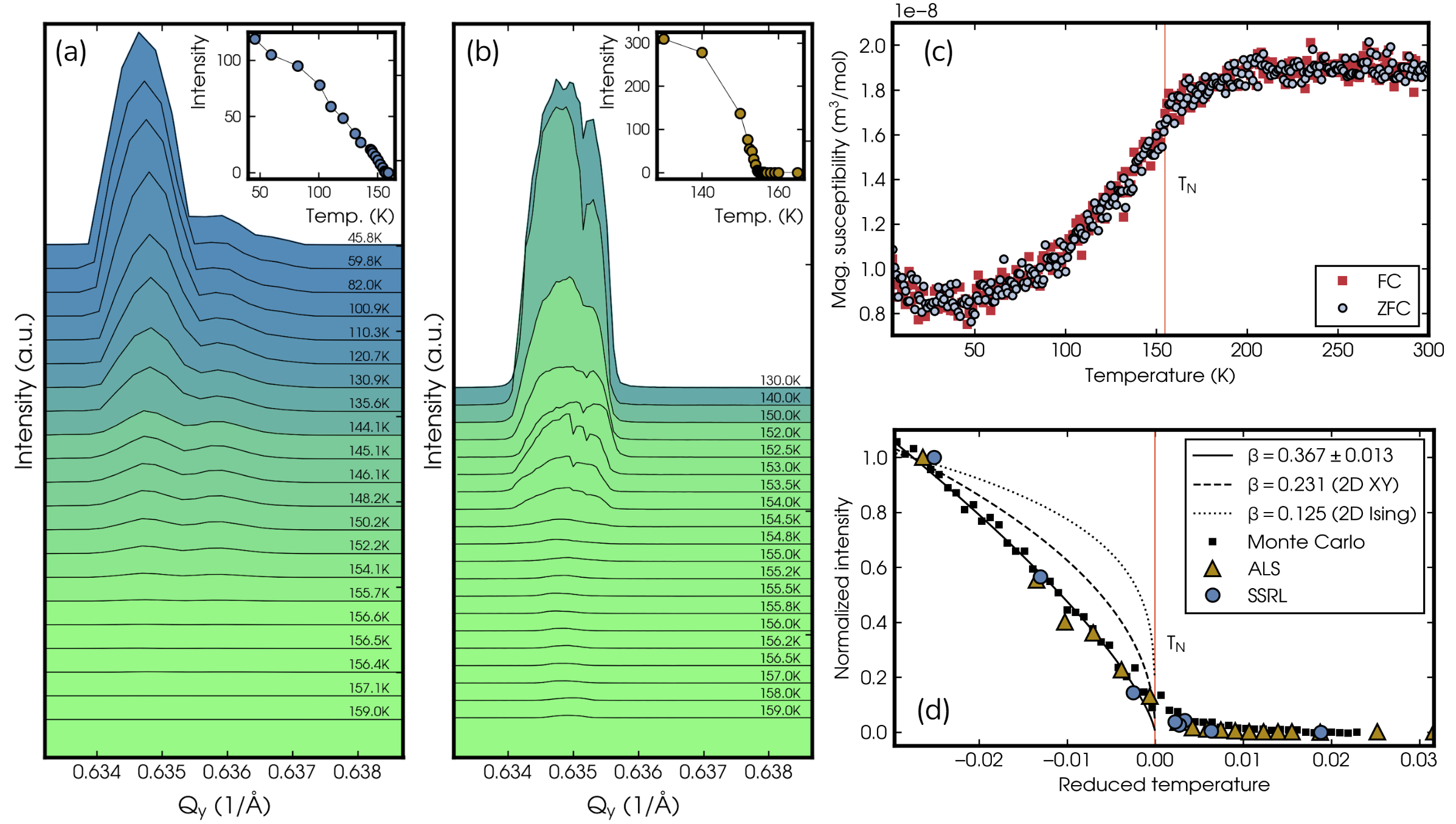}
    \caption{Temperature dependent experiment results.  (a) \& (b) $\theta$-scan intensities of the magnetic $(010)$ Bragg peak collected at 853~eV (Ni $L$-edge) for different temperatures from SSRL and ALS lightsources, respectively.  Intensity traces are scaled and offset for visualization.  Insets show the total intensity at each temperature, normalized to their highest temperature value where the intensity is lowest. (c) Magnetic susceptibility data using vibrating sample magnetometry with an applied $\vb{B}$-field of 1~T $\parallel$ with the $a$-axis for both field-cooled (FC) and zero-field-cooled (ZFC) measurements. (d) Integrated intensity of SSRL (shown in blue) and ALS (shown in orange) normalized to their value at $T=0.97\times T_N$ shows their functional form as they approach the magnetic phase transition. The N\'eel temperature $T_N=155.5$~K is marked with a vertical line.  The curves corresponding to the fitted scaling value $\beta=0.367\pm0.013$ and the theoretical values for the other models for members of this family of $\beta=0.231$ (2D  XY) and $\beta=0.125$ (2D Ising) are shown together with the data.}\label{Fig_Tempdpdt}
\end{figure*}

The atomic structure is shown in Fig.~\ref{Fig_unitcell}a, with the Ni atoms shown in solid red. The $(0k0)$ peaks are forbidden by this crystal structure for odd $k$, but since the `zig-zag' AFM order (one spin chain with spin pointing along $a$, and the opposite spin chain pointing in the $\bar{a}$-direction) possesses a different symmetry than the parent lattice, the magnetic structure can be directly studied at these forbidden reflections without background from non-resonant charge scattering. No scattering was observed at the magnetic reflections off-resonance at base temperature.  Tuning the beam energy to the Ni $L$-edge, an incredibly strong resonance was observed with an energy width of $\sim2$~eV at the same momentum transfer.  Fig.\ref{Fig_Tempdpdt} shows the temperature dependent resonant magnetic scattering intensity from integrated rocking curves of the $(010)$ reflection. The measurements show a decrease in intensity as the sample is warmed until vanishing at the $T_N$ corresponding to the antiferromagnetic/paramagnetic phase transition, which was also confirmed with magnetic susceptibility measurements of the same sample (see Fig.~\ref{Fig_Tempdpdt}c).

The soft X-ray data confirmed a temperature dependence in agreement with the `zig-zag' magnetic order parameter and transition from an AFM to a paramagnetic phase at $T_N$=155.5~K.  The X-ray scattering intensity $I$ -- here the integration of the $\theta$-rocking curve of the $(010)$ reflection -- is proportional to the sublattice magnetization squared, and can be fit to the form, $I \propto -\alpha \tau^{2\beta}$ for the ordered phase ($T<T_N$), where $\tau$ is the reduced temperature $(T-T_N)/T_{N}$ and $\alpha$ is a constant of proportionality. In order to compare the intensities measured at both instruments, both curves are normalized to their value closest to $T = 0.97 \times T_N$, as shown in Fig.\ref{Fig_Tempdpdt}d. Note that critical scaling analysis is only meaningful very near to the transition temperature~\cite{pelissetto2002critical}.  The fitted value $\beta=0.367\pm0.013$ was obtained by fitting both sets of data from the SSRL and ALS measurements together in this region. These dual soft x-ray studies confirmed the magnetic ground state, N\'eel temperature, and yielded the critical index, which we can use to compare our observations with predictions made from theory. Universality of the critical temperature scaling points the 3D Heisenberg universality class, which we will explore further with \textit{ab initio} calculations. 

We perform independent first-principles Density Functional Theory (DFT) calculations, and use these results to initiate additional theoretical calculations using Density-Matrix Renormalization Group (DMRG) and Monte Carlo simulations. We outline the results here while the theoretical details can be found in the supplementary materials. Because there is a large discrepancy in the literature for \nips, we used first-principles DFT \cite{hohenberg1964inhomogeneous} to extract the magnetic exchange parameters using two different methods, designated ``Method A'' and ``Method B'' (as shown in Table \ref{tab:coupling_constants}). DFT was performed here using the Vienna {\it ab-initio} simulation package (VASP)~\cite{kresse1996efficient, kresse1999ultrasoft}. Second, to calculate the magnetic exchange parameters, we used the single-particle Green’s function constructed from a low-energy tight-binding model by the atom-centered Wannier function using Ni-$d$ and $s$-$p$ orbitals in the Wannier90 code suite \cite{marzari1997maximally} (See supplementary materials for details). The exchange parameters are then used to build an atomistic spin model based on the Heisenberg Hamiltonian in Eq.\ref{Eq:HamiltonianJ1J2J3} and simulate the temperature effect on the magnetization following the Monte Carlo Metropolis algorithm \cite{metropolis1953equation} as implemented in the Vampire software package \cite{evans2014atomistic}. This predicted the N\'eel temperature to be $T_N = 153\,K$ and was confirmed to have good agreement with the magnetic susceptibility and X-ray data (Fig.\ref{Fig_Tempdpdt}). The Monte Carlo temperature dependence of the squared magnetization is shown in Fig.\ref{Fig_Tempdpdt}d. Finally, to confirm the zig-zag magnetic ordering in the ground state, the exchange parameters were also used to calculate the ground state using the DMRG method. Due to the $3d^8$ Ni atom having two unpaired electrons in the $e_g$ orbitals, we use the spin-$1$ Heisenberg model with exchanges $J_1$, $J_2$, $J_3$, and $J_4$ on the honeycomb lattice with single-ion anisotropy $\textbf{A}_{ii}$. The definition of this spin-Hamiltonian is,

\begin{eqnarray}
  H= -\sum_{i<j} J_{ij}~\textbf{S}_i\cdot \textbf{S}_j - \sum_{i}\textbf{S}_i \cdot \textbf{A}_{ii} \cdot \textbf{S}_i
  \label{Eq:HamiltonianJ1J2J3}
\end{eqnarray} where the $\textbf{S}_i$ are the quantum spin operators acting on the $i^\text{th}$ site. $J_{ij}$ equals $J_1$ when $i$ and $j$ are the nearest neighbors and equals $J_2$ and $J_3$ for second and third nearest neighbors, respectively.  For the Monte Carlo simulation of the bulk spin dynamics, the small interlayer exchange $J_4$ is included. The DMRG result confirms the stability of the magnetic structure in the single layer and that the magnetic exchange parameters yield a zig-zag order for the ground state measured here, with ferromagnetic chains along the $a$-axis and antiferromagnetic modulation along the $b$-axis with modulation vector $\vb{Q}_{mag} = [010]$ as shown in Fig.\ref{Fig_unitcell}. The spins are canted slightly above and below the $ab$-plane. 

\begin{table}[tb]
\centering 
\begin{tabular}{l|c c c c c}  
\hline\hline 
DFT Method & $J_1$ & $J_2$ & $J_3$ & $J_4$ &$A^{zz}$ \\ [.5ex] 
\hline 
Method A & $2.4852$  & $0.0828$ & $-9.4798$ & $0.0196$ & --\\
Method B & $3.81$ & $0.15$ & $-14.6$ & $0.08$ & $-0.10$\\
\hline\hline 
\end{tabular}
\caption{Magnetic coupling constants (in the unit of meV) calculated using two different DFT methods: Method A, the TB2J software~\cite{he2021tb2j} and Method B, the Four-State Energy Mapping (4SEM)~\cite{TotalEnergyJ} for bulk \nips. A negative $J_3$ value indicates the 3rd-neighbor superexchange is AFM. $A^{zz}$ is the single-ion anisotropy.}\label{tab:coupling_constants}
\end{table}

\section{Discussion}
Although these types of TMCP materials are primarily thought to exhibit 2D magnetism due to their vdW structure, the critical exponent measured here, and confirmed with theory, is surprisingly consistent with the universality class of the 3D Heisenberg model. We have demonstrated that our experimental findings suggested a 3D Heisenberg-like spin Hamiltonian, and that this model predicts a static ground state that is consistent with our data.  To further explore this, the atomistic spin model which was used to calculate the temperature effect on the magnetization via Monte Carlo simulation was used for annealing the magnetic state to simulate the temperature dependence of the order parameter. Fitting these calculated values as a function of temperature also shows the order parameter exponent to be $\beta=0.36$, within our error bars as measured through the soft X-ray measurements. This further confirms that our experimental observations and first-principles calculations using the Heisenberg 3D model yield a consistent picture in \nips. 

The critical exponent closely matches the three-dimensional Heisenberg universality class ($\beta = 0.366$), which is characterized by a three-component order parameter and an $O(3)$ symmetry, usually used to describe isotropic magnets \cite{campostrini-prb-2002, gibertini2019magnetic}. This result is intriguing as it is well above the value for the critical exponent assigned to spin-Hamiltonians from 2D universality classes~\cite{taroni2008universal}. This was determined in the same sample for low temperatures \cite{wildes2015nips3}, and in related systems, i.e. the 2D XY model ($\beta$ = 0.231) for MnPS$_3$ \cite{xy-PhysRevB.74.094422}, or the 2D Ising model ($\beta$ = 0.125) for FePS$_3$~\cite{ising-Nauman_2021}. 

This result shows \nips~can be described by a three-dimensional and three-component spin-1 Heisenberg model and a single-ion anisotropy with a weak interlayer exchange $J_4$. This was hinted at in the neutron diffraction work mentioned above which also analyzed the temperature dependence of the magnetic order parameter near the transition and found a value of $\beta=0.3$, implying a crossover to 3D-like behavior \cite{wildes2015nips3}. While the value of $\beta=0.3$ for high-$T$ puts the spin-Hamiltonian in the 3-dimensional regime, the question of the precise form taken among 3D Ising ($\beta=0.326$), 3D XY ($\beta=0.348$) and 3D Heisenberg ($\beta=0.366$), which are distinguished by their number of relevant spin components ($n=1$, $n=2$, and $n=3$ respectively), was left unanswered. The data fitted from our experiments directly confirm the 3D Heisenberg behavior from the critical exponent. This has been hinted at using other methods, such as a recent study by Scheie et al, which mapped the magnetic Brillouin zone using Inelastic Neutron Scattering (INS) and observed a weak inter-plane magnon dispersion corresponding to a ferromagnetic exchange of -0.38~meV~\cite{scheie2023spin}. Taken together, these studies provide complimentary insights into the magnetic properties of \nips, and share similar qualitative conclusions about the dimensionality of the spin dynamics.

We further find that the AFM Heisenberg exchange, in particular the third-nearest neighbor exchange $J_3$ dominates the Hamiltonian. Though the interlayer ferromagnetic exchange $J_4$ is crucial for introducing the 3D character, it is much smaller in comparison to the in-plane interactions and insufficient for creating long-range correlations along the $c$-axis. Hence, \nips~can be thought of as an easy-plane antiferromagnet, with magnetic vdW layers which are weakly coupled ferromagnetically. Further exploration will be needed to determine an exact description of the Hamiltonian, where details are emerging in the lower energy magnon excitations, such as a possible bi-axial characteristics and other effects that cannot be explained by linear spin wave theory~\cite{scheie2023spin, jana2023magnon}.  New tools are being developed in both the X-ray and neutron scattering frontiers to elucidate these small energetic differences at the fringe of current experimental capabilities~\cite{plumley2023ultrafast, chitturi2023capturing}.

The main motivation here however was to use resonant X-rays to probe the magnetic structure as well as its length scale to provide direct evidence of the AFM state. Measuring the in-plane AFM order using X-rays poses significant challenges due to the inherent limitations of accessing in-plane features of thin vdW crystals, which is typically attempted in a grazing incidence from the plane orthogonal to the $c$-axis direction. However, our study demonstrates that it is indeed feasible to probe the AFM order in \nips~using both soft (and hard) X-rays by scattering resonant X-rays directly from the edge of the layer. We show clean growth is enough to provide a smooth region of the surface to be accessed, without resorting to polishing or cleaving the sample, which would result in either powderization or destroying the material. This work opens up new possibilities for investigating AFM materials using more advanced X-ray scattering techniques, such as pump-probe studies at X-ray free electron lasers \cite{plumley2023ultrafast}, where ultra-fast magnetism can be explored, or fast fluctuations using advanced pulse structure methods \cite{decker-2022-scirep} combined with coherent X-rays, such as in X-ray Photon Fluctuation Spectroscopy \cite{Shen-mrsa-2021}. By successfully measuring the AFM order directly with X-rays, we provide experimental evidence that expands the applicability of X-ray methods to a wide range of low-dimensional magnetic systems.

In conclusion, our study demonstrates the use of resonant X-ray diffraction to measure the AFM order directly along the vdW layer in \nips, using single crystals, opening new avenues for investigation in these types of quasi-2D magnetic systems. The magnetic reflections were strong enough to be observed for two different types of experiments, at soft X-ray energies using the $L$-edge in Ni, where magnetic signal is usually large for transition metals, as well as for higher energies using resonance at the $K$-edge (see supplemental materials). These results were verified at three different X-ray synchrotron light sources. Our observations confirm the `zig-zag' AFM structure previously measured by neutron diffraction, and is further supported by our DMRG ground state calculations for a spin-1 Heisenberg Hamiltonian on a honeycomb lattice. This theoretical calculation was made possible by first-principles DFT, used to extract the parameters needed for the ground state. Additional theoretical evidence was reported in solving the Landau–Lifshitz–Gilbert spin dynamics equation through Monte Carlo simulations, which could reproduce the N\'eel temperature and the critical exponent within the errors of our measurement. Through this experimental and theoretical evidence, we find the vdW antiferromagnet \nips~--conventionally considered a 2D XY magnet -- is unexpectedly more closely described by the 3D Heisenberg universality class. In other words, the magnetism can not be thought of as a two-component order parameter, nor can the interlayer interaction be neglected. It will be interesting to explore this finding further by studying other members of the TMCP family with X-rays, described by different Hamiltonians, to see if the 3D nature is more general or a special case for the largely in-plane spin-1 system. Moreover, because fluctuations dominate the behavior in 2D, it would be enlightening to explore the fluctuations directly of these magnetic vdW systems as has been done in other systems \cite{seaberg-2021-prr}, to see if they map on to the requisite universality classes. 

\section{Methods}
\subsection{Sample preparation}
Single crystals of \nips~were used for the series of experiments in this study. They were grown with optimized growth parameters by \textit{HQ Graphene} using chemical vapor transport. The crystals naturally grow as hexagonal platelets. From a typical batch, the thickest crystals were selected with the most well-defined edges for the resonant x-ray experiments. Energy-dispersive X-ray spectroscopy was then used to confirm the composition and the $a$ and $b$ edges were identified by orienting the crystals in a Bruker D8 Venture single crystal x-ray diffractometer.

\subsection{X-ray beamline experiments}
All experiments were conducted with a single \nips~crystal mounted on a copper sample holder using silver paint. Different single crystal specimens were used in each experiment, however their shape and dimensions are approximately the same.  The thickness of the edge in the $c$-direction is roughly 100~\textmu m and the mass of the crystals are $\sim 2.5$~mg. Mechanically, the crystals are soft and easily deformed. The quality of the edge surface is especially critical for soft X-ray experiments due to the increased surface sensitivity at lower X-ray photon energies.  Additionally, the magnetic characteristics of vdW crystals are known to vary dramatically with lattice strain and stacking arrangement~\cite{mcguire2020cleavable, gibertini2019magnetic}.  With this in mind, the crystals were surveyed with a Scanning Electron Microscope (SEM) at the Stanford Nano Shared Facilities laboratory in order to locate the regions along the edge with minimal surface disorder.

Two resonant X-ray diffraction experiments were carried out using soft X-rays at the Ni $L_3$-edge (853~eV).  The first was at Beamline 13-3 of the Stanford Synchrotron Radiation Light Source (SSRL), and the second was at Beamline 7.0.1.1 of the Advanced Light Source (ALS).  Rresonant X-ray diffraction was also measured at the Ni $K$-edge (8.332~keV) at Beamline 4-ID-D of the Advanced Photon Source (APS) using a four-circle diffractometer. Vibrating Sample Magnetometry (VSM) was used to characterize the temperature dependence of the magnetic susceptibility and confirm AFM magnetic ordering in the sample.  No differences between the field-cooled and zero-field-cooled were observed, and the magnetic susceptibility was in good agreement with previously reported measurements~\cite{wildes2015nips3}.

The soft X-ray data from beamline 13-3 of the SSRL was collected at the Ni L-edge (853~eV) in a vertical scattering geometry as illustrated in Fig. 1.a using a 4-circle diffractometer.  The sample was mounted to a copper puck using colloidal silver paint. The energy was selected using a spherical grating monochromater. The sample was cooled using a liquid helium cryostat and sustained under ultra-high vacuum.  The beam spot size at the sample location was approximately 250 microns.  X-ray scattering intensity was measured using a Hamamatsu GaAsP photodiode.

The soft X-ray data from beamline 7.0.1.1. of the ALS was collected at the Ni L-edge (853~eV) in a vertical scattering geometry as illustrated in Fig. 1.a..  The energy was selected using a spherical grating monochromater. The sample was cooled using a liquid nitrogen cryostat and sustained under high vacuum.  The sample was mounted to a copper puck using colloidal silver paint.  X-ray scattering intensity was measured using a Si photodiode. A pinhole was used to achieve a small beam spot with size approximately 8 microns.  

The hard X-ray data (shown in the supplementary materials) at the Ni K-edge (8.332~keV) was collected in a vertical scattering geometry with horizontal polarization as illustrated in Fig.~1a using a 4-circle diffractometer.  The energy was selected using a double crystal Si (111) monochromator, which was detuned by 2~arcsec in order to reduce the intensity to 87\% of the maximum.  The beam shape was measured to approximately 0.18~mm in the horizontal direction and 0.11~mm in the vertical.  The sample was mounted to a copper puck using colloidal silver paint and was cooled with a liquid helium cryostat. An energy dispersive Silicon drift detector was used to measure the elastic X-ray scattering intensity.  This removes the K$\alpha$ fluorescence but not the K$\beta$, which can be seen in Supplementary Figure~3 as background in the fixed-$\vb{Q}$ energy scan measurements.

\subsection{Density functional theory}
To get insight into the magnetic structure of the $\mathrm{NiPS_3}$, we have performed the first-principle calculations within the framework of the density functional theory (DFT)~\cite{hohenberg1964inhomogeneous} using the Vienna {\it ab-initio} simulation package (VASP)~\cite{kresse1996efficient, kresse1999ultrasoft}. The ground state electronic structure is obtained with the projector augmented-wave pseudo-potential, while the electron exchange-correlation effects are considered by the generalized gradient approximation (GGA)~\cite{perdew1996generalized} with Perdew-Burke-Ernzerhof (PBE) parametrization. The strong correlation effect of the valance Ni-$d$ orbitals is corrected by considering an effective onsite Hubbard potential ($U_{\rm eff}$) ~\cite{hubbard_U, Anisimov_1997}. Throughout our calculations, we choose this value to be $U_{\rm eff}$ = 4.0 eV. The weak interlayer vdW interaction is treated by the DFT-D3 correction of Grimme with zero-damping function~\cite{grimme2010consistent}. The energy cut-off of 350~eV was used for the plane-wave basis set, and the integration of the Brillouin zone was performed with an 11$\times$9$\times$11 $\Gamma$-centered $k-$mesh \cite{Monkhorst1976}. The total energy tolerance criteria are set to $10^{-5}$ eV to satisfy self-consistency. We used the experimental structure parameters while the ionic positions were optimized until the residual forces on each ion were less than 10$^{-2}$ eV/\AA $~$ and the stress tensors became negligible.\\

We used two methods to calculate the magnetic exchange parameters. Firstly, we used the single-particle Green’s function with the rigid spin-rotation as perturbation as proposed by the Liechtenstein, Katsnelson, Antropov, and Gubanov (LKAG) \cite{liechtenstein1987local} and implemented in TB2J code \cite{he2021tb2j}. To obtain the single-particle Green's function, we construct a low-energy tight-binding model by the atom-centered Wannier function using Ni-$d$ and $s$-$p$ orbital in the Wannier90 code suite \cite{marzari1997maximally}. Secondly, we have used the four-state method~\cite{TotalEnergyJ}, where the magnetic interaction atom pair was isolated in a $2\sqrt{2}\times\sqrt{2}\times2$ super-cell. \\  

\subsection{Ground state calculations}
We use the complex number Density-Matrix Renormalization Group (DMRG) method to confirm the magnetic order of the ground state observed by magnetic X-ray diffraction~\cite{White1992}. The Hamiltonian that we use is the spin-$1$ Heisenberg model with $J_1$, $J_2$, and $J_3$ on the single-layered honeycomb lattice with single-ion anisotropy $\textbf{A}_{ii}$ as defined in Equation \eqref{Eq:HamiltonianJ1J2J3}.

The coupling constants that we have obtained from the four-state energy mapping method \cite{TotalEnergyJ} are $J_1=3.81$ meV, $J_2=0.15$ meV, $J_3=-14.6$ meV. The minus sign before $J_3$ means that the superexchange between two spins is antiferromagnetic; otherwise, it is ferromagnetic coupling. 

We also use the four-state energy mapping method to computer the Single-Ion Anisotropy (SIA) and show a non-zero small component for the diagonal part of the SIA Hamiltonian, \textit{i.e.} $A_{ii}^{zz}(S^{z}_i)^2$ with $A_{ii}^{zz}=-0.1$ meV. Practically in DMRG calculation, we use the $J_1$ as the energy unite and redefine $J_{2(3)}/J_1$ and $A_{ii}^{zz}/J_1$ for simplification. The lattice geometry is cylindrical with two basis vectors $\textbf{e}_1 = (\textbf{a}, 0)$ and $\textbf{e}_2 = (\textbf{a}/2, \sqrt{3}\textbf{b}/2)$ for each unit cell, and considering periodic and open boundary conditions in the $\textbf{e}_2$ and $\textbf{e}_1$ directions, respectively, with width $N_y$ and length $N_x$, where $N_y$ and $N_x$ are the numbers of spins along the $\textbf{e}_2$ and $\textbf{e}_1$ directions. Note that each unit cell contains two spins. This paper focuses primarily on the four-leg cylinder with $N_y=4$ and $N_x=12$ because the DMRG method is more precise for computing quasi-one dimensional systems `with finite cylinder widths but a relatively longer cylinder length. By keeping up to $2187$ states in each DMRG sweep with a typical truncation error $\epsilon\sim 10^{-5}$, we obtained a variational ground state with magnetic ordering. The spin symmetry-breaking directions of each spin have three components, $\langle S^x_{i}\rangle$, $\langle S^y_{i}\rangle$ and $\langle S^{z}_i\rangle$. If without SIA, all spins form the zig-zag order, but each spin is canted with more pronounced $\langle S^{x}_i\rangle$ and $\langle S^{y}_i\rangle$ components and the smallest out-of-plane $\langle S^{z}_i\rangle$ component. However, the finite SIA further suppresses the smallest $\langle S^{z}_i\rangle$ component and drives the magnetic ordering to a perfect zig-zag order within the $ab$-plane. Conclusively from our simulation, even without the interlayer coupling, the zig-zag ordered state can be stabilized by the intralayer Heisenberg spin exchange $J_1$, $J_2$ and $J_3$ with ferromagnetic $J_1$ and $J_2$ plus more significant antiferromagnetic $J_3$. The introduction of SIA further suppresses the out of $ab$-plane spin component, allowing the 2D zig-zag order to form. 

\section*{Data Availability}
The data presented in this study will be made available on reasonable request.

\section{Acknowledgements}
Special thanks to Rick Scholtens for his assistance in aligning the orientation of the crystal surface edge after the growth process.  This work was primarily supported by the U.S. Department of Energy (DOE), Office of Science, Basic Energy Sciences under Award No. DE-SC0022216. Portions of this work were also supported by the by the U.S. Department of Energy (DOE), Office of Science, Basic Energy Sciences under Contract DE-AC02-76SF00515 both for the Materials Sciences and Engineering Division and for the Linac Coherent Light Source (LCLS), at the SLAC National Accelerator Laboratory, operated by Stanford University.  J. J. Turner acknowledges support from the U.S. DOE, Office of Science, Basic Energy Sciences through the Early Career Research Program. This research used resources of the National Energy Research Scientific Computing Center, a DOE Office of Science User Facility supported by the Office of Science of the U.S. Department of Energy under Contract No. DE-AC02-05CH11231 using NERSC award BES-ERCAP0023852.  This research used resources of the Advanced Photon Source, a U.S. DOE Office of Science user facility operated for the DOE Office of Science by Argonne National Laboratory under Contract No. DE-AC02-06CH11357. Part of this work was performed at the Stanford Nano Shared Facilities, supported by the National Science Foundation under award ECCS-2026822. The \nips~unit cell image in Figure \ref{Fig_unitcell}.a was created using \textit{VESTA}~\cite{momma2011vesta}.

\section*{Author Contributions}
R.P., T.A., L.S., N.B., Z.P., A.P., A.I., J.S.L., S.M., S.R., G.F., and J.T. collected the data. R.P. analyzed data, with input from L.S, Z.P., A.P., M.D., W.S., A.L., and J.T., S.M., C.P., J.N., H.C., A.F., A.B., S.C. produced the theoretical calculations, modeling, and simulations. R.P., S.M., C.P., J.T. wrote the manuscript. R.P., T.A., E.B., and J.T. conceived of and designed the research. All authors contributed to discussing, reviewing, editing, and approving the final manuscript.

\section*{Competing Interests}
The authors declare no competing interests.

\section{Supplementary Material for: 3D Heisenberg universality in the Van der Waals antiferromagnet \nips}
\subsection{Supplementary Note: Additional hard X-ray characterization at the Ni K-edge.}

The $(0k0)$-scans for $k=1,2,5,6$ measured at the Ni $K$-edge at $T=100$~K are shown in Fig.\ref{Fig_Braggpeaks}. The odd being magnetic and even structural, with the (010) being the strongest magnetic peak measured. The magnetic structure was confirmed by scanning the upstream monochromater to vary the incident X-ray photon energy at constant-$Q$, where resonant magnetic enhancement at 8.332~keV was observed at the specific magnetic ordering vectors (see Fig.\ref{Fig_Energyscans}).  This enhancement was surprisingly large for the Ni pre-edge, up to a factor of 10 for the lowest order magnetic reflection, which is usually attributed to a quadrupolar transition as $1s$ electrons are directly promoted to $3d$ orbitals, but this can also be due to the dipole channel with hybridization~\cite{hill1997k, wang2021defect}.

We checked azimuthal conditions, with both the $a$- and $c$-axes orthogonal to the scattering plane to confirm that the spin-moments are oriented largely along the $a$-axis as reported by neutron scattering measurements \cite{wildes2015nips3}. Furthermore, the temperature dependence of this (010) reflection was measured to confirm the scattering intensity was magnetic in origin, as it disappeared at the N\'eel temperature. The width of the Bragg peaks shown in Fig.\ref{Fig_Braggpeaks} can be used to estimate a lower bound for the correlation length, which was found to be $\xi_{\parallel}>3000$~\AA~for both the structural and magnetic reflections at 100~K. This indicates that the long-range, in-plane magnetic order is limited by the crystallinity of the atomic parent lattice.  The peak is very broad along the direction corresponding to the out-of-plane $c$-axis, as is commonly observed in layered crystals.  The lower bound of the out-of-plane correlation length was estimated to be $\xi_{\perp}>100$~\AA.

\begin{figure}[h!]
    \begin{center}
    \hspace{-0.75cm}
    \includegraphics[width=0.45\textwidth]{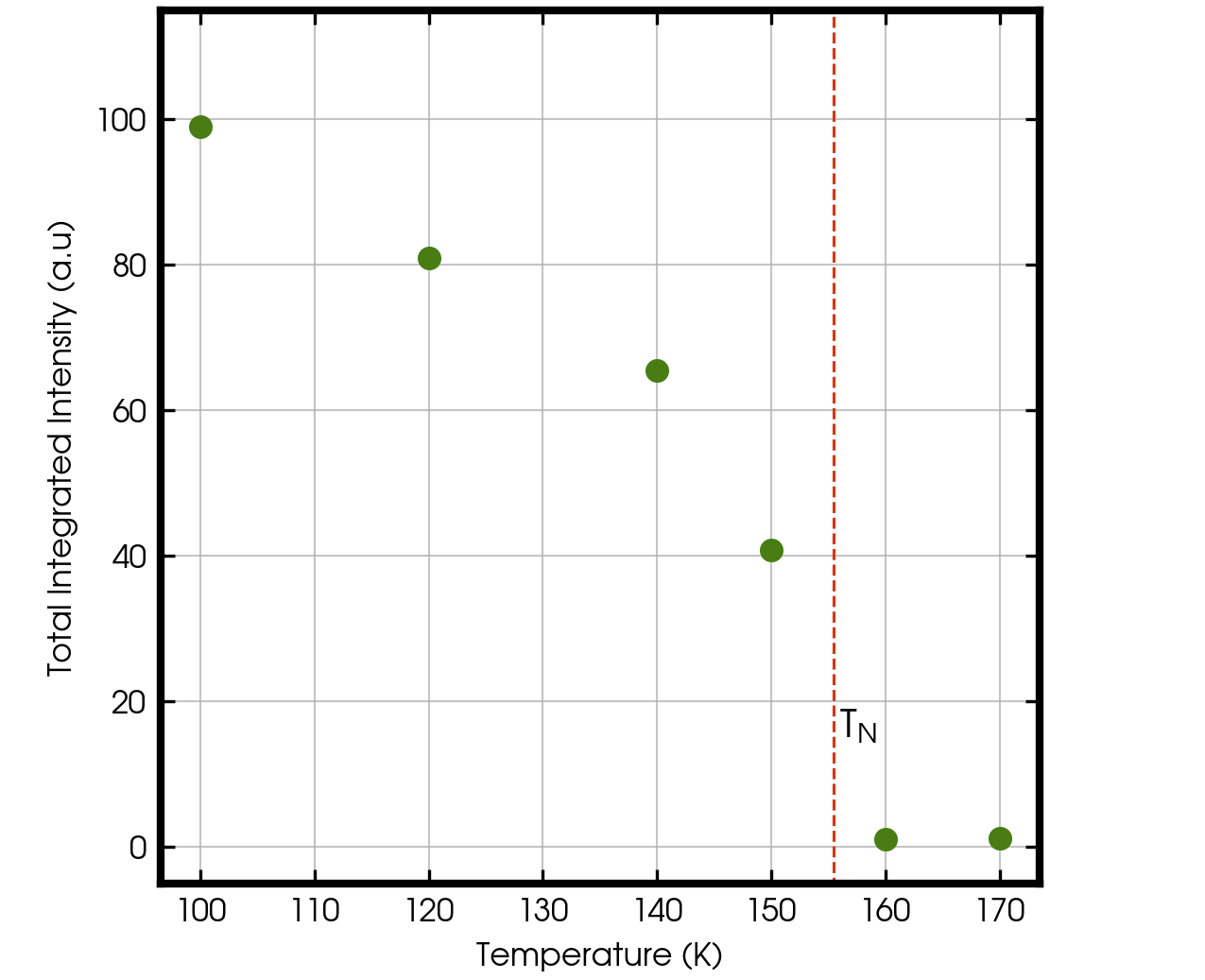}
    \caption{Integration of the $\theta$ rocking curve intensity from hard X-ray data collected at Beamline 4-ID-D of the APS.  The signal was derived from the (010) reflection at the $K$-edge and disappeared at the N\'eel temperature of $T=155.5$~K}\label{Fig_HXRtempdep}
    \end{center}
\end{figure}

\begin{figure}
    \begin{center}
    \hspace{-0.75cm}
    \includegraphics[width=0.45\textwidth]{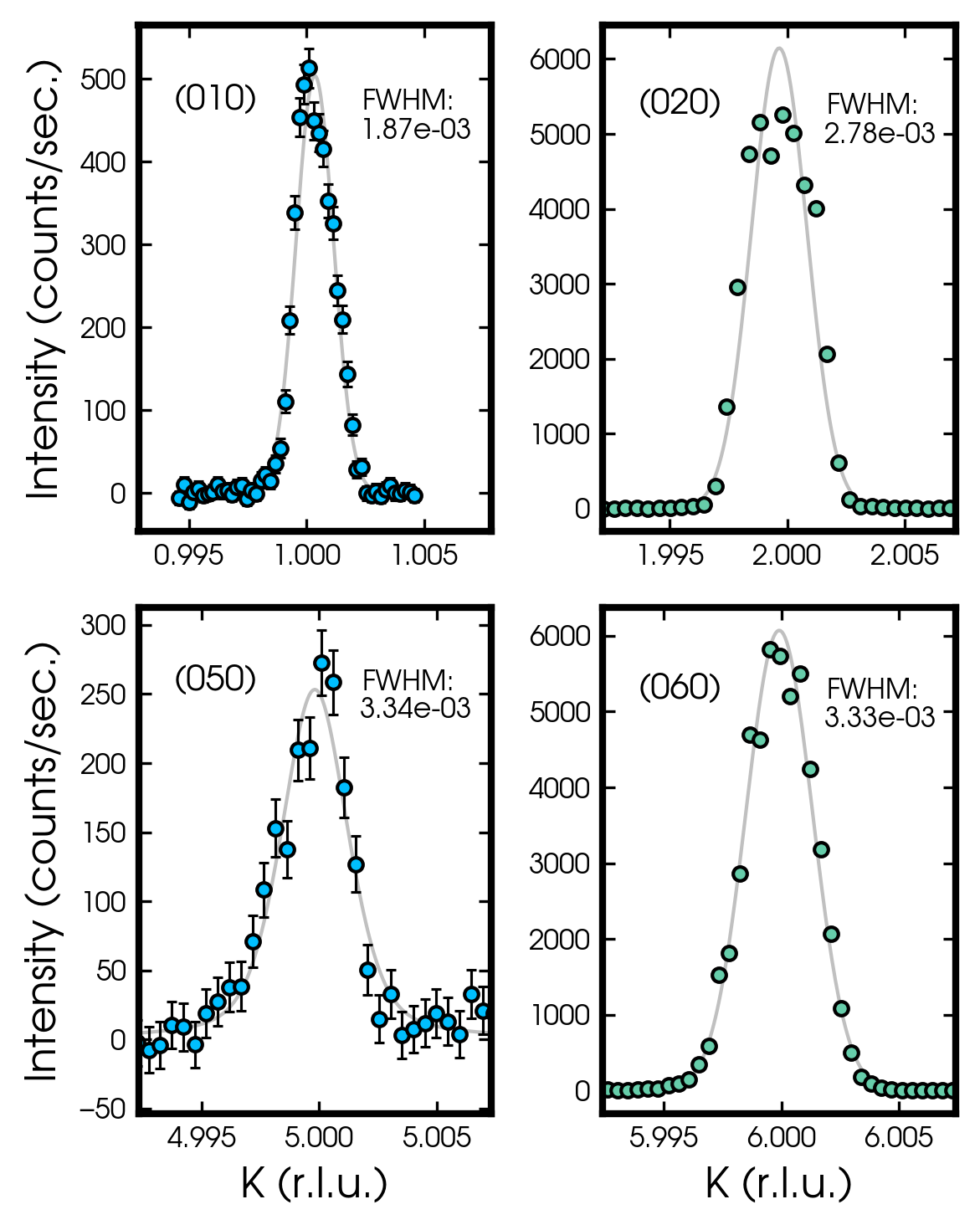}
    \caption{Magnetic (left) and structural (right) Bragg peaks measured at the APS beamline 4-ID-D.  All peaks were measured at $T=100$~K with $E_\gamma = 8.332$~keV (Ni $K$-edge) and fit using pseudo-Voigt models to obtain their FWHM.  Background from fluorescence was subtracted so that the value at the tails is zero.}\label{Fig_Braggpeaks}
    \end{center}
\end{figure}

\begin{figure}
    \includegraphics[width=0.45\textwidth]{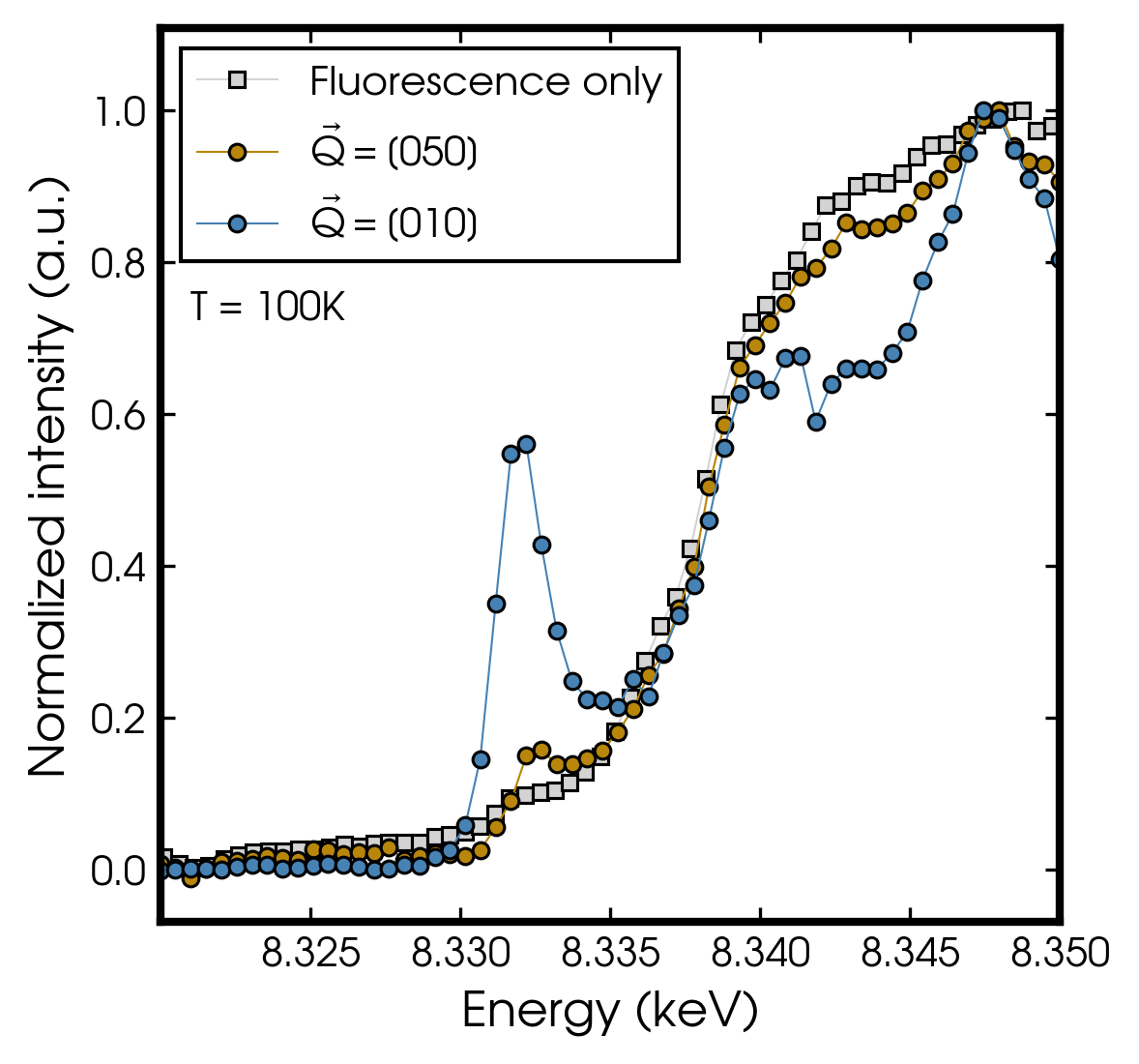}
    \caption{Energy scans (fixed $\vb{Q}$).  Fluorescence only measurements were taken with the detector position perpendicular to the scattering plane. Each curve is normalized to their maximum value.  A strong enhancement occurs at the pre-edge energy when the magnetic Bragg condition is met.}\label{Fig_Energyscans}
\end{figure}
\bibliographystyle{unsrt}
\bibliography{bib}

\end{document}